\documentclass[aps,10pt]{revtex4}

\usepackage{psfrag}
\usepackage{subfigure}
\usepackage{color}
\usepackage{mathrsfs}
\usepackage{graphicx}
\usepackage[colorlinks=true,linkcolor=blue,citecolor=magenta,urlcolor=blue]{hyperref}
\usepackage{amssymb, bm}
\usepackage{amsmath, amsthm}
\usepackage{epstopdf}
\usepackage{hyperref}
\usepackage{enumerate}
\usepackage{longtable}
\usepackage{float}
\usepackage{array, multirow}
\usepackage{fancyhdr}
\usepackage{amsfonts} 

\usepackage{mathrsfs}
\usepackage{amssymb}
\newcommand{\Ce}{\widetilde{C}}

\begin{document}	
	\title{A family of non-autonomous hybrid Li\'enard oscillators based on a parametrically extended commutative factorization}

	\author{J. de la Cruz}
	\email{josue.delacruz@ipicyt.edu.mx; ORCID: 0000-0001-5943-5752}
	\affiliation{Instituto Potosino de Investigaci\'on Cient\'{\i}fica y Tecnol\'ogica,\\ 
		Camino a la Presa San Jos\'e 2055, Colonia Lomas 4a Secci\'on,
		78216 San Luis Potos\'{\i}, S.L.P., Mexico}

\author{H.C. Rosu}
	\email{hcr@ipicyt.edu.mx; ORCID: 0000-0001-5909-1945; Corresponding author} 
	\affiliation{Instituto Potosino de Investigaci\'on Cient\'{\i}fica y Tecnol\'ogica,\\ 
		Camino a la Presa San Jos\'e 2055, Colonia Lomas 4a Secci\'on,
		78216 San Luis Potos\'{\i}, S.L.P., Mexico}

    \author{G. Gonz\'alez}
    \email{gabriel.gonzalez@academicos.udg.mx; ORCID: 0000-0001-8217-321x}
    \affiliation{Departamento de Ciencias B\'asicas y Aplicadas, Universidad de Guadalajara, CUTonal\'a Avenida Nuevo Perif\'erico 555, Ejido San Jos\'e Tateposco, 
    45425, Tonal\'a, Mexico}
    
    \author{O. Cornejo-P\'erez}
    \email{octavio.cornejo@uaq.mx; ORCID: 0000-0002-1790-7640} 
    \affiliation{Facultad de Ingenier\'{\i}a, Universidad Aut\'onoma de Quer\'etaro, Centro Universitario Cerro de las Campanas, 76010
     Santiago de Quer\'etaro, Mexico}

	\bigskip
	\bigskip
 \begin{abstract}
 	We introduce a class of nonautonomous nonlinear oscillator equations of mixed Li\'{e}nard type
	that arises from a parametric deformation of the commutative factorization procedure applied to second-order ordinary
	differential equations with periodic solutions. Their solutions, in particular the isochronous waveforms, are obtained in closed form through a Riccati reduction scheme for
	the power-law choice of the factorization function, $\phi(x)=kx^{q}$, $k\in \mathbb{R}$, and $q\in\mathbb{N}$, corresponding to the so-called modified Emden oscillators, 
    and do not depend on the arbitrary deformation parameter $A_1$.  
	The variational structure of the equation is characterised by a Lagrangian
	supplemented with a generalised Rayleigh dissipation function that contains
	a non-standard cubic term in $\dot{x}$.
	We also show that multiplying the equation of motion by the Jacobi multiplier
	$M(x)=x^{-2A_1}$ a position-dependent-mass (PDM) form is obtained with related friction and restoring force.
  
		\medskip
		\end{abstract}
	\maketitle
 
Isochronous nonlinear oscillators
attracted considerable interest since Calogero's systematic study of the phenomenon \cite{r0}, and Li\'enard-type equations $\ddot x+f(x)\dot x+g(x)=0$ have proved a particularly fertile setting for constructing them. Two decades ago, Chandrasekar, Senthilvelan and Lakshmanan showed that the cubic oscillator $\ddot x+kx\dot x+(k^2/9)x^3+\omega^2x=0$ shares the exact period of the linear harmonic oscillator ($k=0$) regardless of amplitude \cite{r4}, a result subsequently extended to arbitrary-dimensional and coupled Li\'enard-type systems \cite{r13} and to schemes for generating $N$-dimensional isochronous nonsingular Hamiltonian systems \cite{r12}. Closer to the hybrid class-I/class-II structure treated in the present paper, the mixed Li\'enard-type equation $\ddot x+f(x)\dot x^2+g(x)\dot x+h(x)=0$ -- containing both a term linear and a term quadratic in $\dot x$, exactly as in our Eq.~\eqref{eq:9} below -- has been analyzed from the inverse-problem and symmetry perspective, with its isochronous solutions identified through the interplay of Lie point symmetries and the Jacobi last multiplier \cite{r15}. In parallel, Guha and collaborators developed the Jacobi last multiplier as a systematic route to Lagrangian and Hamiltonian structures for Li\'enard-type systems and their isochronicity \cite{r10}, a program extended very recently to the Levinson--Smith equation $\ddot z+J(z)\dot z^2+F(z)\dot z+G(z)=0$ -- structurally the same hybrid form treated here -- which was shown to map to the harmonic oscillator via a nonlocal transformation and to possess a bi-Hamiltonian structure amenable to canonical quantization \cite{r11,r14}.

Second-order ordinary differential equations (ODE) with periodic solutions are conveniently factored in the form \cite{r1,r2,r3}
\begin{equation}\label{eq:1}
	\left(D_t+\phi(x)+\frac{i}{\Ce}\right)\left(D_t+\phi(x)-\frac{i}{\Ce}\right)x=0, \qquad D_t= \frac{d}{dt}~,
\end{equation}
where $\Ce\in\mathbb{R}$, and $\phi(x)$ is a corresponding factorization function. For example, the factoring $(D_t+i\omega)(D_t-i\omega)x=0$ of the harmonic oscillator equation is obtained by taking $\phi=0$ and $\Ce=1/\omega$.

However, the same equation can be considered in a more general framework by defining the function
\begin{equation}\label{eq:2}
	\Phi(x,t):=\left({\cal D}_{xt}+\phi(x)-\frac{i}{\Ce}\right)x~,   
\qquad {\cal D}_{xt}:=\frac{dx}{dt}\frac{\partial}{\partial x}+\frac{\partial}{\partial t}
\end{equation}
implying 
\begin{equation}\label{eq:2b}
\left({\cal D}_{xt}+\phi(x)+\frac{i}{\Ce}\right)\Phi(x,t)=0
\end{equation}
as the equivalent first order PDE of \eqref{eq:1}. If in \eqref{eq:2b} one assumes the separable form $\Phi(x,t)=\varsigma(t)\,x$,
one obtains from \eqref{eq:2b} and \eqref{eq:2} the Bernoulli equation
\begin{equation}\label{eq:3}
	\frac{d\varsigma}{dt} + \frac{2i}{\Ce}\varsigma + \varsigma^2=0~,
	\end{equation}
whose solution is
\begin{equation}\label{eq:4}
	\varsigma(t) = -\frac1\Ce\left[i+\tan\left(\frac{t+\delta}{\Ce}\right)\right]~.
\end{equation}
This Bernoulli solution plays an important role in the method of commutative factorization \cite{r1,r2,r2b} as it yields to nonlinear isochronism\cite{r6}.

In this paper, within the commutative factorization approach, we define a non-autonomous dynamical system for which the role of the Bernoulli equation \eqref{eq:3} is taken by a Riccati equation, and show that for a particular form of its free term, the non-autonomous system has the same periodic solutions as the autonomous one. We apply this scheme to the modified Emden oscillators with monomial factorization functions \cite{r4}.

The paper is organized as follows. In Section~2, we develop the extended factorization scheme for a general factorization function $\phi(x)$ and present its Rayleigh Lagrangian and position-dependent mass (PDM) formulations.  
In Section~3, we apply the results of Section~2 to the case of the cubic modified Emden oscillator corresponding to $\phi=kx$ and extend the monomial factorization function to higher orders, $\phi=kx^q$, for any $q\in \mathbb{Z}^+$. The general closed-form solution valid for every $q$ and every $A_1$ (with the detailed derivation in the Appendix) is presented and the periodicity criterion separating odd and even $q$ is formulated. The conclusions and additional discussion are included in Section 4.

\section*{2. THE $A_1$-EXTENDED COMMUTATIVE FACTORIZATION}

We consider the following generalization of Eq.~\eqref{eq:1}:
\begin{equation}
	\left(\frac{d}{dt}+\phi(x)+\frac{i}{\Ce}\right)\left(\frac{d}{dt}+\phi(x)-\frac{i}{\Ce}\right)x=\alpha(t)\,x + A_0\Phi + A_1\frac{\Phi^2}{x}~.
	\label{eq:5}
\end{equation}
Substituting $\Phi=\varsigma(t)\,x$, we obtain the Riccati equation
\begin{equation}
	\frac{d\varsigma}{dt} + \frac{2i}{\Ce}\varsigma + \varsigma^2 = \alpha(t) + A_0\varsigma + A_1\varsigma^2~,
	\label{eq:6}
\end{equation}
and by grouping terms,
\begin{equation}
	\frac{d\varsigma}{dt} + 2\left(\frac{i}{\Ce}-\frac{A_0}{2}\right)\varsigma + (1-A_1)\varsigma^2 = \alpha(t).
	\label{eq:7}
\end{equation}
Choosing $A_0=2iA_1/\Ce$, this Riccati equation takes the simplified form
\begin{equation}
	\frac{d\varsigma}{dt} + (1-A_1)\left[\frac{2i}{\Ce}\varsigma+\varsigma^2\right] = \alpha(t)~.
	\label{eq:8}
\end{equation}
If we further choose the free term as
\[
\alpha(t) = -\frac{A_1}{\Ce^2}\sec^2\left(\frac{t+\delta}{\Ce}\right)~,
\]
then its solution coincides with the Bernoulli solution \eqref{eq:4} for arbitrary $A_1$~.

Therefore, we conclude that the non-autonomous system with periodic solutions has the form

\begin{equation}\label{eq:9}
\ddot{x} + \left[2(1-A_1)\phi + x\frac{d\phi}{dx}\right]\dot{x} - \frac{A_1}{x}(\dot{x})^2 + \left[(1-A_1)\phi^2 
+\frac{1}{\Ce^2}\left( 1 + A_1\tan^2\left(\frac{t+\delta}{\Ce}\right)\right)\right]x = 0~.
\end{equation}
where $\dot{x}=dx/dt$. This is our working equation in the following, which is generically valid for arbitrary factorization function $\phi(x)$.


\subsection*{2.1 Lagrangian and PDM formulations}

The nonlinear ODE \eqref{eq:9} can be obtained from the equation of motion of a dissipative Lagrangian representation \cite{Mustafa2021,Mustafa2023, Rayleigh}
\begin{equation}\label{eq:L}
	L(x,\dot{x},t)
	= \frac{1}{2}\dot{x}^{2}
	- (1-A_1)\int^{x}\!\phi^{2}(\xi)\,\xi\,d\xi
	- \frac{1+A_1\tan^{2}\!\left(\frac{t+\delta}{\Ce}\right)}{2\Ce^{2}}\!x^{2}~,
\end{equation}
where the integral encodes the nonlinear stiffness generated by $\phi(x)$. Because the equation of motion
contains a term quadratic in $\dot{x}$, which cannot be derived from a standard
quadratic Rayleigh function, one must supplement $L$ with the generalised dissipation function
\begin{equation}\label{eq:R}
	\mathcal{R}(x,\dot{x})
	= \frac{1}{2}\!\left[2(1-A_1)\phi(x)+x\phi'(x)\right]\dot{x}^{2}
	- \frac{A_1}{3x}\,\dot{x}^{3}~,
\end{equation}
where $\phi'(x)=d\phi/dx$ and the time derivative of the cubic contribution satisfies $\partial_{\dot{x}}(-A_1\dot{x}^{3}/3x) = -(A_1/x)\dot{x}^{2}$.

The equation of motion follows from $\tfrac{d}{dt}\partial_{\dot{x}}L - \partial_{x}L + \partial_{\dot{x}}\mathcal{R} = 0$.
Since $L$ is quadratic in $\dot{x}$ with unit coefficient, the time-derivative term contributes simply $\ddot{x}$. The potential term in $L$ yields
\[
-\frac{\partial L}{\partial x}
= \frac{x}{\Ce^{2}}\!\left[(1-A_1)\Ce^{2}\phi^{2}(x)+1+A_1\tan^{2}\!\left(\frac{t+\delta}{\Ce}\right)\right],
\]
while differentiating $\mathcal{R}$ with respect to $\dot{x}$ produces the
velocity-dependent forces,
\[
\frac{\partial\mathcal{R}}{\partial\dot{x}}
= \Bigl[2(1-A_1)\phi(x)+x\phi'(x)\Bigr]\dot{x} - \frac{A_1}{x}\,\dot{x}^{2}~.
\]
Hence \eqref{eq:9} is obtained, which is a Li\'{e}nard equation with both a linear and a quadratic velocity term \cite{Mustafa2023,Ruiz}.
The quadratic term $-(A_1/x)\dot{x}^{2}$ is non-conservative and therefore cannot
originate from a standard kinetic energy, $\mathcal{R}$ must contain
the non-classical cubic term whose $\dot{x}$-derivative reproduces it.

The classical PDM formulation \cite{Sara,Oscar} is related to a Li\'enard II equation. In order to obtain a similar form we multiply \eqref{eq:9} throughout by an
integrating factor $\mu(x) = M(x)$, to be determined, such that the result takes the form
\begin{equation}\label{eq:PDM_derived}
	M(x)\,\ddot{x}	+ \tfrac{1}{2}M'(x)\,\dot{x}^{2}+ \Gamma(x)\,\dot{x}+ V'(x)= 0~.
\end{equation}
This PDM equation is not the classical one, due the friction term $\Gamma(x)$; this version has been studied in \cite{Ruiz}. Comparing the $\dot{x}^{2}$ coefficients between the multiplied equation \eqref{eq:9} and
\eqref{eq:PDM_derived} requires
\[
-M(x)\,\frac{A_1}{x} = \frac{1}{2}M'(x)~,
\]
which is the separable ODE
\begin{equation}
	\frac{M'(x)}{M(x)} = -\frac{2A_1}{x}~.
	\label{eq:jacobi_ode}
\end{equation}
Integrating \eqref{eq:jacobi_ode} gives $\ln M(x) = -2A_1\ln x + \text{const}$,
and setting the integration constant to zero without loss of generality yields
\begin{equation}
	M(x) = x^{-2A_1}~.
	\label{eq:M}
\end{equation}
This is the unique (up to a constant scale factor)
integrating factor that absorbs the quadratic velocity term into the PDM inertia.
The same conclusion follows from the general theory of Li\'{e}nard-type equations
\cite{GhoseChoudhury2015,Chanda2018}
\begin{equation}
	\ddot{x} + f(x)\,\dot{x}^{2} + g(x)\,\dot{x} + h(x,t) = 0,
	\label{eq:lienard_gen}
\end{equation}
for which the integrating factor that absorbs the quadratic velocity term
into the PDM inertia is
\begin{equation}
	\mu(x) = \exp\!\left(2\int f(x)\,dx\right).
	\label{eq:mu_gen}
\end{equation}
With $f(x) = -A_1/x$ one recovers
\[
\mu(x) = \exp\!\left(-2A_1\int\frac{dx}{x}\right) = x^{-2A_1},
\]
consistent with \eqref{eq:M}.

Multiplying \eqref{eq:9} by $M(x)=x^{-2A_1}$, the first two terms become
$x^{-2A_1}\ddot{x} - (A_1 x^{-2A_1}/x)\dot{x}^{2}$.
Since $M'(x) = -2A_1\,x^{-2A_1-1}$, one has $\tfrac{1}{2}M'(x) = -A_1\,x^{-2A_1-1}$,
so the quadratic velocity term is exactly $\tfrac{1}{2}M'(x)\dot{x}^{2}$ and the
pair combines into the PDM inertia $M\ddot{x}+\tfrac{1}{2}M'\dot{x}^{2}$, which is
what the Euler--Lagrange equation for $\tfrac{1}{2}M(x)\dot{x}^{2}$
produces. Multiplication by $M(x)$ also yields to
\begin{align}
	\Gamma(x) &= x^{-2A_1}\Bigl[2(1-A_1)\phi(x)+x\phi'(x)\Bigr]~,
	\label{eq:Gamma}\\
	V'(x)     &= \frac{x^{1-2A_1}}{\Ce^{2}}\!\left[(1-A_1)\Ce^{2}\phi^{2}(x)
	+1+A_1\tan^{2}((t+\delta)/\Ce)\right]~,
	\label{eq:Vprime}
\end{align}
so the multiplied equation is exactly \eqref{eq:PDM_derived}, describing the dynamics of a particle
with a Jacobi position-dependent mass $M(x)$ subject to friction $-\Gamma(x)\dot{x}$ and
restoring force $-V'(x)$. This follows from the PDM Lagrangian
$L_\text{PDM} = \tfrac{1}{2}M(x)\dot{x}^{2} - V(x)$
supplemented by the non-conservative force $-\Gamma(x)\dot{x}$.

The algebraic identity which contains the correspondence is
\begin{equation}
	x^{-2A_1}\frac{\partial}{\partial\dot{x}}\!\left(-\frac{A_1}{3x}\,\dot{x}^{3}\right)
	= -\frac{A_1\,x^{-2A_1}}{x}\,\dot{x}^{2}
	= \tfrac{1}{2}M'(x)\,\dot{x}^{2},
	\label{eq:pivot}
\end{equation}
which shows that the cubic term in $\mathcal{R}$, after multiplication by the Jacobi
multiplier $x^{-2A_1}$, is precisely the velocity-derivative of the inertial
correction $\tfrac{1}{2}M'(x)\dot{x}^{2}$ that a variable mass generates.
Both formulations yield the same equations of motion, since they generate identical equations of motion:
the Rayleigh picture contains a non-classical dissipation function, while the PDM picture absorbs the 
quadratic damping into the non-trivial mass $M(x) = x^{-2A_1}$ and recovers the classical linear-friction 
non-conservative force $-\Gamma(x)\dot{x}$.

\begin{figure}[ht]
	\centering
	\subfigure{\includegraphics[width=0.475\textwidth]{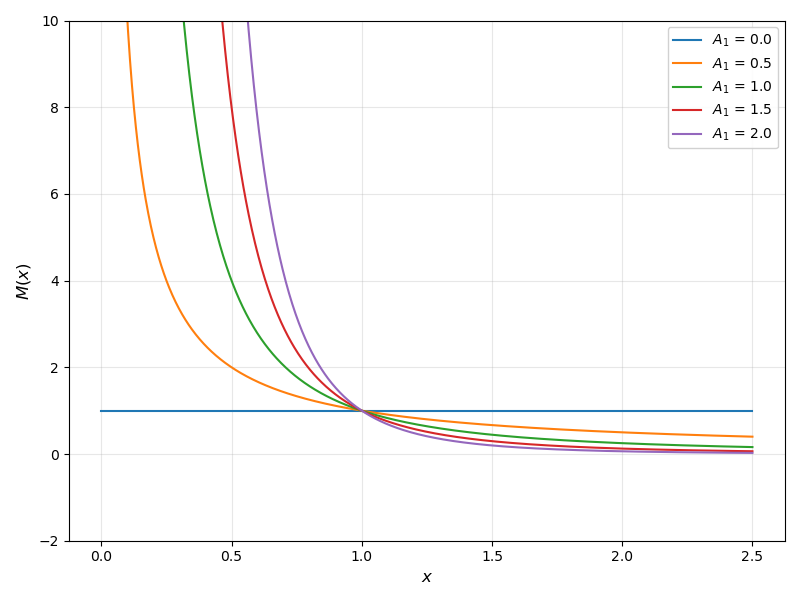}}
	\subfigure{\includegraphics[width=0.475\textwidth]{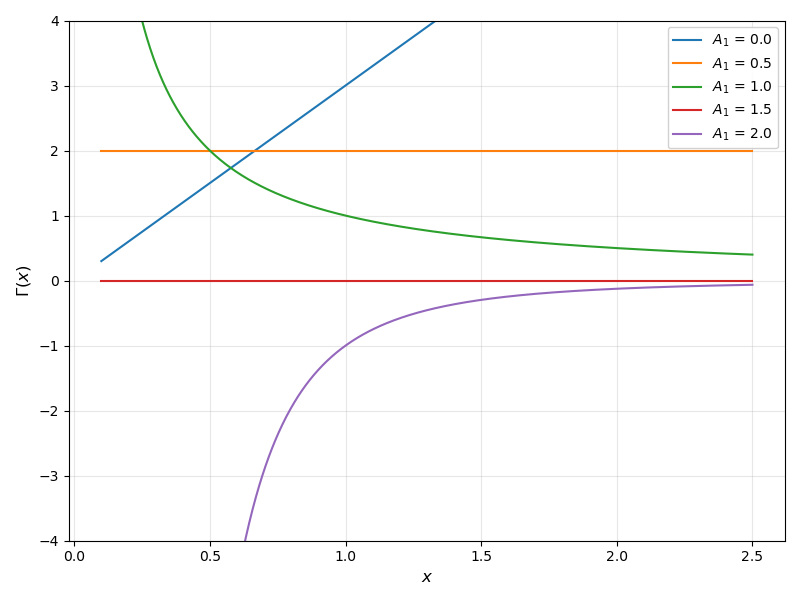}}
	\caption{Left: the effective mass $M(x)=x^{-2A_1}$ for several values of $A_1$.
		Right: the friction coefficient $\Gamma(x)$ for $\phi=kx$ ($\Ce=k=1$, $\delta=0$), which vanishes identically at $A_1^{\star}=3/2$.}
	\label{fig7}
\end{figure}
\section*{3. The $A_1$-FAMILY OF MODIFIED EMDEN OSCILLATORS}

An interesting application is the case of the linear monomial factorization function, $\phi=kx$, turning Eq.~\eqref{eq:9} into
\begin{equation}\label{eq:10}
	\ddot{x} +  (3-2A_1)\,kx\,\dot{x} - \frac{A_1}{x}(\dot{x})^2 + (1-A_1)k^2x^3 + \frac{1+A_1\tan^2\left(\frac{t+\delta}{\Ce}\right)}{\Ce^2}x=0~,
\end{equation}
which describes the motion of a one-parameter family of cubic non-autonomous modified Emden oscillators of mixed Liénard classes, unless the cases $A_1=0$ when it becomes autonomous of Liénard class I and $A_1=3/2$ when it is non-autonomous 
of Li\'enard class II.

The Lagrangian of these oscillators reads
\begin{equation}\label{eq:L_kx}
	L(x,\dot{x},t) = \frac{1}{2}\dot{x}^2 
	- \frac{(1-A_1)k^2}{4\Ce^2}\,x^4
	- \frac{x^2}{2\Ce^2}\left(1 + A_1\tan^2\left(\frac{t+\delta}{\Ce}\right)\right)~.
\end{equation}
with dissipation function
\begin{equation}\label{eq:R_kx}
	\mathcal{R}(x,\dot{x}) = \frac{(3-2A_1)kx}{2}\,\dot{x}^2 - \frac{A_1}{3x}\,\dot{x}^3.
\end{equation}
The general waveform solution of Eq.~\eqref{eq:10}, as shown in the Appendix, is
\begin{equation}\label{eq:11}
	x(t) = \frac{\cos\left(\frac{t+\delta}{\Ce}\right)}{c_1+k\Ce\sin\left(\frac{t+\delta}{\Ce}\right)}~,
\end{equation}
where $c_1$ is an integration constant. It does not depend on the parameter $A_1$, and is identical to the general solution of the cubic autonomous modified Emden oscillator ($A_1=0$) \cite{r4}.

\begin{figure}[h]
	\centering
	\includegraphics[width=\textwidth]{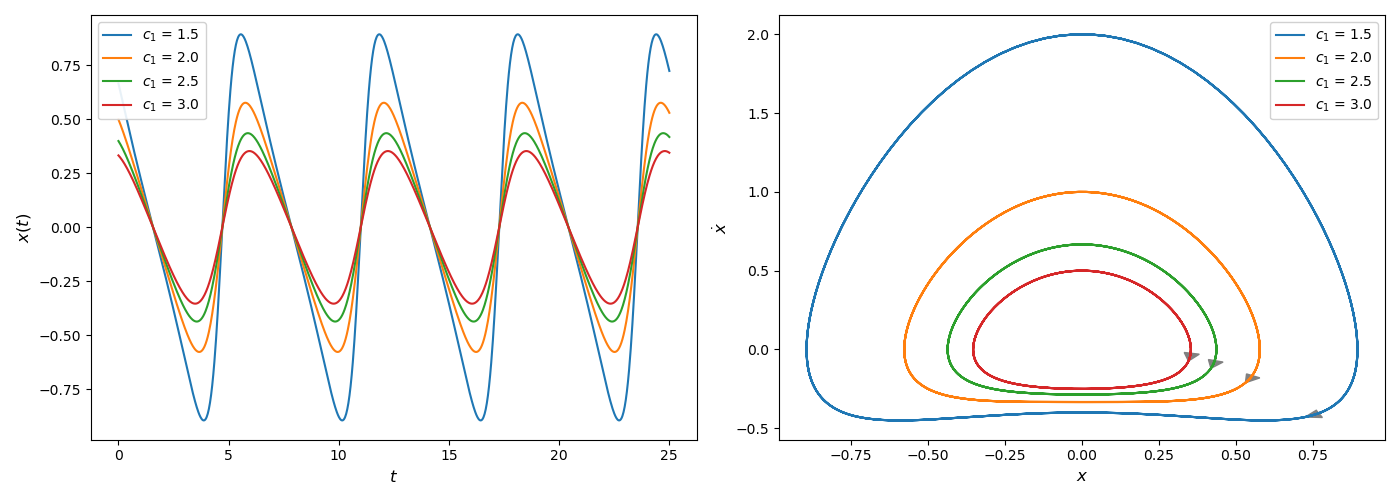}
	\caption{The isochronous waveform solutions \eqref{eq:11} and its phase portraits for $k=\Ce=1$, $\delta=0$, and various values of the integration constant $c_1$. These waveforms are the same for any value of the deformation parameter $A_1$.}
	\label{fig1}
\end{figure}

This solution has constant period $T=2\pi\Ce$, i.e.\ it is isochronous and nonsingular if and only if
\begin{equation}\label{eq:12}
	\left|\frac{c_1}{k\Ce}\right| > 1~.
\end{equation}
Some plots and phase portraits of the isochronous waveforms, Eq.~\eqref{eq:11}, are shown in Fig.~\ref{fig1}.
They can be also considered as amplitude-modulated deformations of pure harmonic signals, $x(t)=A(t)\cos((t+\delta)/\Ce)$ with $A(t)=1/(c_1+k\Ce \sin ((t+\delta)/\Ce))$.

\subsection*{3.1 Higher-order monomial factorization function $\phi=kx^q$, $q\in\mathbb{Z}^+$}

Starting from the generalized ODE \eqref{eq:9} with $\phi=kx^q$ leads to:
\begin{equation}\label{eq:24}
	\ddot{x} + k(2-2A_1+q)\,x^q \dot{x} - \frac{A_1}{x}\dot{x}^2 + (1-A_1)k^2x^{2q+1} + \frac{1}{\Ce^2}\left(1+A_1\tan^2\left(\frac{t+\delta}{\Ce}\right)\right)x = 0~.
\end{equation}

\begin{figure}[ht]
	\centering
	\subfigure{\includegraphics[width=0.475\textwidth]{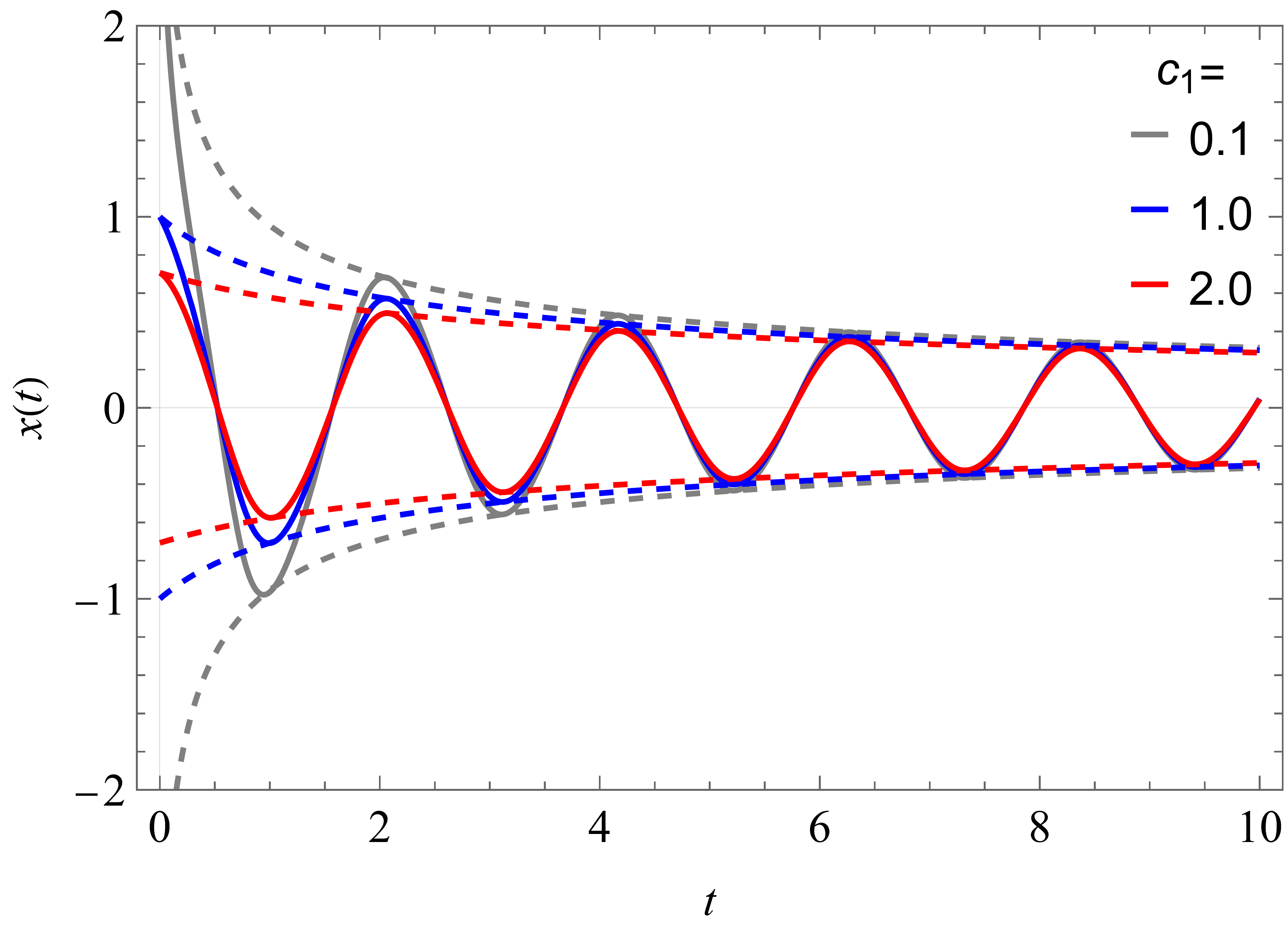}}
	\subfigure{\includegraphics[width=0.475\textwidth]{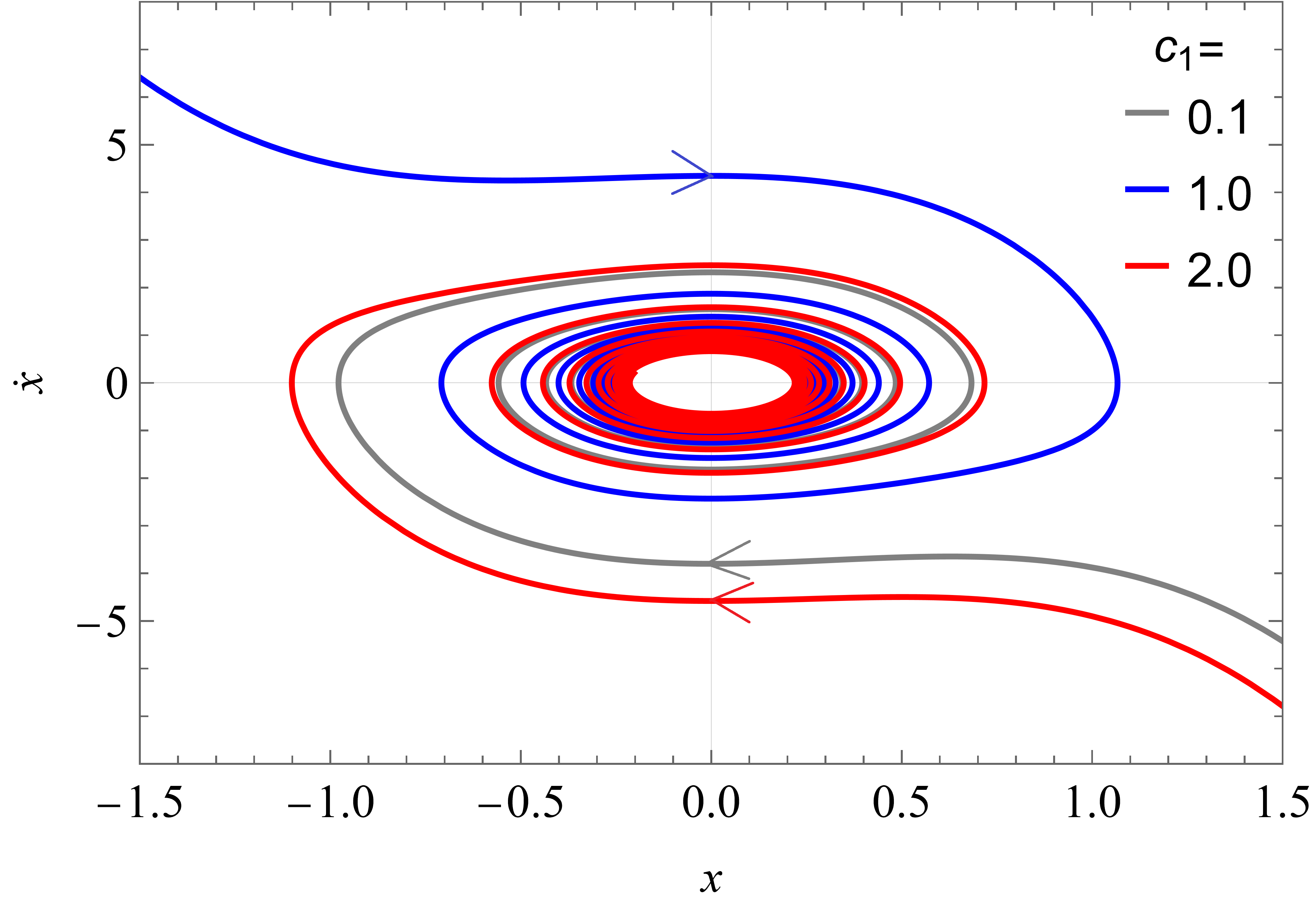}}
	\caption{Left: Solutions $x(t)$ for $q=2$, $\Ce=k=1$, $\delta=0$, and various initial conditions $c_1$, valid for any deformation parameter $A_1$. The singularities for these cases are located on the negative time semiaxe. Right: The corresponding phase portraits.}
	\label{fig-q2}
\end{figure}

This ODE can be also obtained as generalized Euler-Lagrange equation of motion from the dissipative Rayleigh Lagrangian 

\begin{equation}\label{eq:L_kxq}
	L(x,\dot{x},t) = \frac{1}{2}\dot{x}^2 
	- \frac{(1-A_1)k^2}{2(q+1)\Ce^2}\,x^{2q+2}
	- \frac{x^2}{2\Ce^2}\left(1 + A_1\tan^2\left(\frac{t+\delta}{\Ce}\right)\right)~,
\end{equation}
with Rayleigh dissipation function \eqref{eq:R} 
\begin{equation}\label{eq:R_kxq}
	\mathcal{R}(x,\dot{x}) = \frac{k\!\left(2(1-A_1)+q\right)}{2}\,x^q\,\dot{x}^2 
	- \frac{A_1}{3x}\,\dot{x}^3~.
\end{equation}
The higher-order general waveform solution, 
\begin{equation}\label{eq:25}
	x(t) = \frac{\cos\left(\frac{t+\delta}{\Ce}\right)}{\Big[c_1 + qk\displaystyle\int^t\cos^q \left(\frac{s+\delta}{\Ce}\right)\,ds\Big]^{1/q}}~,
\end{equation}
is obtained via the Riccati reduction scheme as detailed in the Appendix. 
In particular, for $q=1$, this reproduces the solution given in Eq.~\eqref{eq:11}. 
The periodicity of these waveforms presents two branches because of the odd-even splitting under selected $q$. To see this, we
denote $\Theta=(t+\delta)/\Ce$ and ${\cal I}(\Theta)=qk\Ce\int_0^\Theta\cos^q s\,ds$. The symmetry $\cos^q(\Theta+\pi)=(-1)^q\cos^q\Theta$ implies:
\begin{itemize}
	\item \textbf{$q$ odd:} $\cos^q$ is anti-periodic under $\Theta\to\Theta+\pi$, so its average over a period vanishes; ${\cal I}(\Theta)$ is bounded and oscillatory. The solution \eqref{eq:25} is globally regular and $2\pi$-periodic in $\Theta$ if and only if \cite{r2}
	\begin{equation}
		|c_1| \;>\;2^{\left(\frac{q-1}{2}\right)}\left[\frac{\left(\frac{q-1}{2}\right)!}{q!!} \right]q|k\Ce|\, ~.  
	\end{equation}
For $q=1$, this condition reduces to \eqref{eq:12}.
	\item \textbf{$q$ even:} $\cos^q$ is 
	$\pi$-periodic with strictly positive mean $\frac1\pi\int_0^\pi\cos^qs\,ds=2^{-q}\binom{q}{q/2}$ (Wallis average), so ${\cal I}(\Theta)$ grows secularly and changes sign at $t\to\pm\infty$. No choice of $c_1$ avoids a singularity for even $q$. These kind of singular solutions are illustrated in Fig.~\ref{fig-q2} for the $q=2$ case together with their phase portraits. Away from the singularity, these waveforms show an underdamped behaviour with a power-law decaying envelope.
\end{itemize}

\section*{4. CONCLUSIONS AND DISCUSSION}

A parametrically deformed family of nonautonomous nonlinear oscillator equations of mixed Li\'{e}nard-type I/II arising from an extended commutative factorization procedure, and illustrated with the case of the modified Emden oscillators, has been introduced in this work. When the deformation parameter ($A_1$, in our notation) is naught, the equation is undeformed, autonomous Li\'enard-type I. Moreover the entire family has solutions independent of $A_1$, which suggests that the deformation parameter can be used as a tuning parameter as well.

As equations of motion, the members of this family admit two equivalent variational descriptions underlined by the identity $x^{-2A_1}\partial_{\dot{x}}(-A_1\dot{x}^{3}/3x)=M'(x)\dot{x}^{2}/2$ that show two different features of the same dynamics.
In the Rayleigh formulation, the unit-mass Lagrangian is supplemented by a generalised dissipation function containing a cubic term $-(A_1/3x)\dot{x}^{3}$ due to the deformation parameter $A_1$, alongside the standard quadratic term.
This cubic contribution is non-standard from the classical Rayleigh dissipation standpoint, but it is precisely required to reproduce the quadratic velocity damping $-(A_1/x)\dot{x}^{2}$ through $\partial_{\dot{x}}\mathcal{R}$.
In the PDM formulation, that same term is reabsorbed into the kinetic part via the Jacobi multiplier $M(x)=x^{-2A_1}$, which is 
the unique solution of the first-order ODE $M'/M=-2A_1/x$ obtained by matching the $\dot{x}^{2}$ coefficients after multiplication.
The resulting PDM equation of motion has the classical structure of a variable-mass particle subject to a $\dot{x}$-friction non-conservative force $-\Gamma(x)\dot{x}$, at the cost of a non-trivial inertia $M(x)=x^{-2A_1}$.

The friction coefficient $\Gamma(x)=x^{-2A_1}[2(1-A_1)\phi+x\phi']$
vanishes identically at $A_1^{\star}=3/2$ for $\phi=kx$ ($A_1^{\star}=(q+2)/2q$ for $\phi=kx^q$), a
distinguished value at which the motion of the PDM particle becomes purely inertial in the standard interpretation; an autonomous oscillator of this kind is briefly mentioned in \cite{GonzalezYakhno2023}. 
On the other hand, in the dissipative Rayleigh function framework for the $A_1$-deformed nonlinear equations, the non-conservative sector relies only on the nonstandard cubic Rayleigh function $\mathcal{R}^\star=-A_1^\star\dot{x}^{3}/3x$, which is a non-conservative gain contribution that interacts with the non-autonomous part to change the dynamics to a purely inertial regime. This is a dynamical behaviour which is possible only for equations of the $A_1$-modified Emden type as one can infer from $A_1=1+x\phi'/2\phi$ demanding for $x\phi'/\phi$ be constant. 

To this end, as a first task for future work on these lines, it would be interesting to study this kind of deformation for other factorization functions, beyond the monomial case presented here. 
\section*{Acknowledgements}

The first author acknowledges the financial support of SECIHTI through a postdoctoral
fellowship.

\appendix
\section*{Appendix}

General $q$.
We substitute 
\[
 x = u^{-1/q}~, \qquad \dot{x} = -\frac{1}{q}\,u^{-(q+1)/q}\dot{u}~,
\qquad \ddot{x} = \frac{q+1}{q^{2}}\,u^{-(2q+1)/q}\dot{u}^{2}
- \frac{1}{q}\,u^{-(q+1)/q}\ddot{u}
\]
into Eq.~\eqref{eq:24} and multiply through by $-q^{2}u^{(2q+1)/q}$ which gives
\begin{equation}
	qu\ddot{u}
	+ \bigl[A_1-(q+1)\bigr]\dot{u}^{2}
	+ kq(q+2-2A_1)\dot{u}
	- q^{2}(1-A_1)k^{2}
	- \frac{q^{2}u^{2}}{\Ce^{2}}\!\left(1+A_1\tan^{2}\!\left(\frac{t+\delta}{\Ce}\right)\right) = 0.
	\label{eq:33theta}
\end{equation}
Dividing by $qu$, one obtains
\begin{equation}
	\ddot{u}
	+ \frac{A_1-q-1}{q}\,\frac{\dot{u}^{2}}{u}
	+ k(q+2-2A_1)\,\frac{\dot{u}}{u}
	- \frac{q(1-A_1)k^{2}}{u}
	- \frac{qu}{\Ce^{2}}\!\left(1+A_1\tan^{2}\!\left(\frac{t+\delta}{\Ce}\right)\right)
	= 0~,
	\label{eq:gen_u}
\end{equation}
in which we use the second substitution
$\varpi = (\dot{u}-c)/u$ with an as-yet undetermined constant $c$ providing
\begin{equation}\label{eq:uc}
\dot{u} = u\varpi + c~, \qquad \ddot{u} = u\dot{\varpi} + u\varpi^{2} + c\varpi,
\qquad (\dot{u})^{2} = u^{2}\varpi^{2} + 2cu\varpi + c^{2}~.
\end{equation}

Substituting \eqref{eq:uc} into \eqref{eq:gen_u} yields to
\begin{eqnarray*}
	\dot{\varpi}u + u\varpi^{2} + c\varpi
	+ \frac{A_1-q-1}{q}\,\frac{u^{2}\varpi^{2}+2cu\varpi+c^{2}}{u}
	+ k(q+2-2A_1)\,\frac{(u\varpi+c)}{u}+&\\
	- \frac{q(1-A_1)k^{2}}{u}
	- \frac{qu}{\Ce^{2}}\!\left(1+A_1\tan^{2}\!\left(\frac{t+\delta}{\Ce}\right)\right)
	= 0&
\end{eqnarray*}
and grouping the terms by powers of $u$, one obtains their coefficients as follows.

$\bullet$ For the $u$ terms:

\begin{equation} \label{eq:33}
	\dot{\varpi} + \frac{A_1-1}{q}\varpi^{2}
	- \frac{q}{\Ce^{2}}\!\left(1+A_1\tan^{2}\!\left(\frac{t+\delta}{\Ce}\right)\right)~.
\end{equation}

$\bullet$ For the $u^{0}$ terms:
\[
c\varpi + \frac{A_1-q-1}{q}\cdot 2c\varpi + k(q+2-2A_1)\varpi
= \varpi\big[q+2(1-A_1)\big]\cdot\frac{qk - c}{q}.
\]

$\bullet$ For the $u^{-1}$ terms:
\[
\frac{A_1-q-1}{q}\,c^{2}
+ k(q+2-2A_1)\,c - q(1-A_1)k^{2}=\big[qc-(1-A_1)(qk - c)\big]\cdot\frac{qk - c}{q}~.
\]

Since the $u^{0}$ and $u^{-1}$ coefficients 
have the common factor $(qk-c)/q$, they vanish identically 
when
\begin{equation}\label{eq:c}
	c = qk~.
\end{equation}

Therefore, under condition \eqref{eq:c}, one can reduce equation \eqref{eq:24} to the Riccati equation provided by the coefficient of the $u$ power:
\begin{equation*}\label{eq:36}
	\dot{\varpi} + \frac{A_1-1}{q}\,\varpi^{2}= \frac{q}{\Ce^{2}}\!\left(1+A_1\tan^{2}\!\left(\frac{t+\delta}{\Ce}\right)\right)~,
\end{equation*}
solved 
by the solution
\begin{equation*}
	\varpi_{p} = \frac{q}{\Ce}\tan\left(\frac{t+\delta}{\Ce}\right)~.
\end{equation*}

Now, in the ODE connecting $\varpi$ and $u$:
\begin{equation*}
	\dot{u} = u\varpi + qk~.
\end{equation*}
one substitutes $\varpi_{p}$:
\begin{equation*}
	\dot{u} - \frac{q}{\Ce}\tan\Theta\;u = qk
	\label{eq:37}
\end{equation*}
and uses the integrating factor:
\[
\mu(t)
= e^{-\frac{q}{\Ce}\int\tan\!\left(\frac{t+\delta}{\Ce}\right)dt}
= \cos^{q}\!\left(\frac{t+\delta}{\Ce}\right)
\]
to obtain:
\[
D_t\!\left[u\cos^{q}\!\left(\tfrac{t+\delta}{\Ce}\right)\right]
= qk\cos^{q}\!\left(\tfrac{t+\delta}{\Ce}\right)~.
\]

Integrating both sides:
\begin{equation*}
	u\cos^{q}\!\left(\frac{t+\delta}{\Ce}\right)
	= qk\int^{t}\!\cos^{q}\!\left(\frac{\tau+\delta}{\Ce}\right)d\tau + c_{1}~.
	\label{eq:38}
\end{equation*}

This leads to $x(t)$ in Eq.~\eqref{eq:25} since $x = u^{-1/q}$.

For the case $q=1$, we obtain 
\begin{equation*}
	u(t)=\sec\!\left(\frac{t+\delta}{\Ce}\right)
	\left[k\Ce\sin\left(\frac{t+\delta}{\Ce}\right) + c_{1}\right]~,
\end{equation*}
from where the solution \eqref{eq:9} results as $1/u(t)$.

\end{document}